\documentclass[aps,prl,twocolumn,nofootinbib,superscriptaddress]{revtex4}
\usepackage[hypertex]{hyperref}
\usepackage{graphicx} % Include figure files
\usepackage{amsbsy}   % Include \boldsymbol command
\usepackage{times}

\def\mysection#1{{\bf #1.}}

\def\beq{\begin{eqnarray}}
  \def\eeq{\end{eqnarray}}
\def\ea{{\it et al.}}
\begin{document}

% Page numbers bottom-center
\pagestyle{plain}

\title{%\begin{flushright}{\small
  %%\vspace*{-1.5cm} LBNL-57250, UCB-PTH-05/04}
  %%\end{flushright}\vspace*{.1cm} 
  Electroweak Baryogenesis from Late Neutrino Masses}

\author{Lawrence J. Hall}
\affiliation{Theoretical Physics Group, 
Ernest Orlando Lawrence Berkeley National Laboratory,
Berkeley, CA 94720}
\affiliation{Department of Physics, University of California,
Berkeley, CA 94720}

\author{Hitoshi Murayama}
\affiliation{Theoretical Physics Group, 
Ernest Orlando Lawrence Berkeley National Laboratory,
Berkeley, CA 94720}
\affiliation{Department of Physics, University of California,
Berkeley, CA 94720}

\author{Gilad Perez}
\affiliation{Theoretical Physics Group, 
Ernest Orlando Lawrence Berkeley National Laboratory,
Berkeley, CA 94720}

%\date{\today}

\begin{abstract}
  Electroweak Baryogenesis, given a first order phase transition, does
  not work in the standard model because the quark Yukawa matrices are
  too hierarchical.  On the other hand, the neutrino mass matrix is
  apparently not hierarchical.  In models with neutrino mass
  generation at low scales, the neutrino Yukawa couplings lead to
  large CP-violation in the reflection probability of heavy leptons by
  the expanding Higgs bubble wall, and can generate the observed
  baryon asymmetry of the universe.  The mechanism predicts new
  vector-like leptons below the TeV scale and sizable $\mu \rightarrow
  e$ processes.
\end{abstract} \pacs{Who cares?} \maketitle

%%%%%%%%%%%%%%%%%%%%%%%%%%%%%%%%%%%%%%%%%%%%%%%%%%%%%%%%%%%%%%%%%%%%%%%%%%%

The neutrino physics has been undergoing a revolutionary progress in the
past several years.  The recent observations of neutrino flavor
conversion in solar, atmospheric and reactor neutrino experiments
provided firm evidence for physics beyond the Standard Model (SM) (see
{\it e.g.}~\cite{PDG,Rev} and refs. therein).  The new data on the
distance/energy-dependence of atmospheric neutrinos~\cite{SKatmnu04}
and the recent spectrum analysis from the reactor
experiments~\cite{Ter} indicate oscillatory behaviour. This strongly
favors the presence of tiny but non-zero neutrino masses.

The most popular explanation for the origin of neutrino masses is the
seesaw mechanism~\cite{seesaw}.  In its minimal form, right-handed
neutrinos are introduced with lepton-number violating Majorana mass
terms, $\Lambda_{\rm L} NN$, as well as Yukawa couplings to the
SM lepton doublets $l$ and the Higgs $h$, $Y_{L} h
Nl$.  The oscillation data requires $\Lambda_{\rm L}\sim 10^{14}~{\rm
  GeV}$ if $Y_L \sim O(1)$.  Moreover, the seesaw mechanism naturally
provides a way to account for the observed baryon asymmetry of the
universe (BAU)~\cite{PDG,WMAP}, $n_B/s\sim8\times 10^{-11}$, where
$n_B/s$ is the baryon to entropy ratio.  The out-of-equilibrium decay
of the right-handed neutrinos creates a lepton asymmetry, which is
partially converted to a baryon asymmetry by the electroweak sphaleron
process (leptogenesis)~\cite{Lep}.  Thermal leptogenesis typically
requires, however, $\Lambda_{\rm L}\sim 10^{9}\,{\rm
  GeV}$~\cite{Thermal} with somewhat small Yukawa couplings $Y_L$ and
large hierarchies in the right-handed masses.

While the seesaw mechanism is very appealing theoretically, it is
unlikely that it will be subject to a direct experimental test in the
near future.  It is important to explore other possibilities for the
origin of neutrino masses.  One such example is the late neutrino mass
framework that induces small neutrino masses due to a low scale of
symmetry breaking~\cite{Glob1}.  When this symmetry is broken, say by
a set of symmetry breaking VEVs, $\langle\phi\rangle=f$, the neutrinos
acquire masses from operators
\begin{equation}
  \left( \frac{\phi}{M_F} \right)^n l N h \;\; (\mbox{Dirac})
  \;\;\; \mbox{or} \;\;\;
  \left( \frac{\phi}{M_F} \right)^{n} \frac{(l h)^2}{\Lambda_L} \;\;
  (\mbox{Majorana}).
  \label{eq:operators}
\end{equation}
We stress that this does not depend on the details of the model, or
whether one uses global~\cite{Glob1} or gauge~\cite{AG,Take,Gauge}
symmetries.  The strongest limits on $f$ arise from big bang
nucleosynthesis~\cite{Glob1,Glob2,Take,Hall,Gauge} and from observation of
supernova neutrinos~\cite{GPS}: for $n=1$, $f \gtrsim 10 \, {\rm keV}$, 
while more powerful limits apply to higher $n$. It is
remarkable that new physics at such a low scale is not excluded by
direct experimental data.

In this letter we show that the late neutrino mass framework can
naturally realize leptogenesis at the electroweak (EW) phase
transition~\cite{foot}.  Heavy vector-like leptons that give rise to
the operators~(\ref{eq:operators}) bounce off the expanding bubble
walls with $O(1)$ Yukawa couplings, and acquire a large asymmetry.
They quickly decay to the standard-model leptons and the sphaleron
process partially converts their asymmetry to the baryon asymmetry.
Thus our mechanism leads to various predictions that can be tested by
near future experiments.  Here we focus on the qualitative features of
our scenario while a more detailed analysis will be presented in a
following publication.  We simply assume that the phase transition is
first order, and focus on the size of the baryon asymmetry as well as
phenomenological constraints on the model.  We briefly comment on the
origin of the first-order phase transition towards the end of this
paper.

%%%%%%%%%%%%%%%%%%%%%%%%%%%%%%%%%%%%%%%%%%%%%%%%%%%%%%%%%%%%%%%%%%%%%%%%%%%%%%%%%
\mysection{The Model and Mechanism} As well as introducing a low scale
$f$, theories of late neutrino masses introduce flavor scales that are
much lower than the scale $\Lambda_L$ of the seesaw mechanism. One
economical possibility, that we explore in this letter, is that these
flavor scales are all of order the EW scale: $M_F, \Lambda_L \sim v
\equiv\langle h\rangle$.  In this case the non-SM states that generate
the operators of~(\ref{eq:operators}) have masses of order $v$ and are
available to take part in EW baryogenesis. The flavor symmetry
breaking scale becomes $f \sim v (m_\nu/v)^{1/n} \sim (m_\nu, \,100
\,\mbox{keV},\, 30 \,\mbox{MeV} ...)$ for $n = 1,2,3 ...$ The $n=1$
case is excluded by BBN and supernova constraints. Large $n$ theories
may be preferred in the sense that the scale $f$ grows and requires
less protection. In this letter, we describe a simple $n=2$ theory
that illustrates our baryogenesis mechanism.  We focus on the
Majorana case since, as shown below, the Dirac case is disfavored by
the direct experimental data for the parameter range that produce
enough baryon asymmetry.

The minimal model for late Majorana neutrino masses is
\begin{eqnarray}
  {\cal L}_\nu=Y \; h N L + M \; L^c L+y \; \phi L^c l
  +\frac{M_N}{2} \; N N +{\rm h.c.}\,, \label{L} 
\end{eqnarray}
where $L,L^c$ are vector-like lepton doublets, and the couplings $Y$,
$y$, and the masses $M$, $M_N$ are 3$\times$3 flavor matrices.  We
consider all eigenvalues of $M$ to be comparable, the same for $M_N$,
and those of $Y$ to be of $O(1)$. This is suggested by the lack of
hierarchy (anarchy~\cite{Hall:1999sn}) in the neutrino masses.  We
refer to their eigenvalues as $\bar{M}$,$\bar{M}_N$, and $\bar{Y}$.
There may be a moderate hierarchy in the eigenvalues of $y$: $y_i$.
Note that there is a a leptonic analogue of the rephasing invariant
Jarlskog determinant even for just two generations
($N_G=2$)~\cite{Maj},
\begin{equation}
  J_L = \Im m {\rm Tr}\left(Y Y^\dagger M_N^* Y^* Y^T M_N
    M_N^* M_N \right).
\end{equation}

Below the scales $\bar{M}$ and $\bar{M}_N$, neutrino masses are
described by the operator $(y_i^2 \bar{Y}^2 / \bar{M}^2 \bar{M}_N)  
l^2 h^2 \phi^2$, corresponding to the $n=2$ case of~(\ref{L}), so 
that the flavor symmetry breaking scale is predicted to be   
\begin{eqnarray}
f^2 \sim {m_{\nu i} v \over  y_i^2 \bar{Y}^2} \;
 {\bar{M}^2 \over v^2} \, {\bar{M}_N \over v}
 \sim { (100~\mbox{keV})^2 \over  y_i^2 \bar{Y}^2} \;
 {\bar{M}^2 \over v^2} \, {\bar{M}_N \over v}.
\label{flavscale}
\end{eqnarray}
A low $f \sim 100~\mbox{keV}$ would imply that the $\phi$ states
contribute to the energy density of the universe during BBN. However,
our baryogenesis mechanism will require $\bar{M}, \bar{M}_N$ somewhat
larger than $v$, and lepton flavor violation will constrain $y_i$ to
be somewhat small.  Hence we expect that the $\phi$ states are heavier
than 1 MeV and decay to neutrinos before BBN.
%A similar result 
%applies for the Dirac case with $n=2$. Such a Dirac theory
%may be generated by introducing a second set of heavy vector leptons
%and restoring lepton number: $M_N = 0$. Below the mass of the heavy 
%leptons the operator $y_i^2 \bar{Y} lN h (\phi^2 /\bar{M}^2)$ leads to
%(\ref{flavscale}) with the $(\bar{M}_N/v)$ factor absent and 
%$\bar{Y}^2$ replaced by $\bar{Y}$.

Let us now describe our main mechanism. With $\bar{M}\sim v$, during
the electroweak phase transition $L$ particles and their CP
conjugates, $\bar L$, are reflected differently from the Higgs bubble
wall. This is due to the presence of unsuppressed CP violating phases
in $Y$, $M$, and $M_N$~(for a related idea see~\cite{Maj,BNV}).  Thus
an asymmetry in $L$ is induced in the region just in front of the
wall. As shown below, the size of this asymmetry is expected to be of
order $J_L/\bar{M}_N^4$.
%% $J_{L}$
%% %= {\rm tr}\left\{Y Y^\dagger M^\dagger \left[M M^\dagger,Y^*
%% %Y^T\right]M\right\}$,  
We assume the $L\to N h$ decay process is kinematically forbidden
({\it i.e.}\/, $\bar{M}_N>\bar{M}$) so that the asymmetry is
transferred to SM leptons via $L\to l\phi$ decays.  The presence of
the asymmetry in the SM lepton doublets biases the sphaleron rate to
induce a ${\rm B}$ production in the vicinity of the wall.  When the
expanding bubble passes over this region, the sphaleron processes
decouple, freezing in a ${\rm B}$ asymmetry.  Outside the bubble the
sphaleron rate $\Gamma_{\rm Sp}\sim \alpha_W^4 T_c$, $T_c$ being the
critical temperature, is much slower than other dynamical scale near
the bubble wall~\cite{CKNRev}.  Thus the baryon asymmetry could be as
large as
\begin{equation}
  \frac{n_B}{s} \sim \frac{1}{g_*} \alpha_W^4 \frac{J_L}{\bar{M}_N^4}
  \sim 10^{-8}
\end{equation}
where $g_* \simeq 100$ is the number of relativistic degrees of
freedom, and anarchical neutrino masses suggest that $J_L/\bar{M}_N^4$
is of order unity.  Below we study what other factors might suppress
the baryon asymmetry.

\mysection{Estimating the Baryon Asymmetry}
To have a semi-quantitative estimation of the resultant BAU, we
apply the thin wall approximation (the validity of this approximation
depends on the details of the mechanism which produces the 1st order
phase transition~\cite{CKNRev}).  We first estimate the
density difference between $L$ and $\bar L$, $n_{L}$,
induced by the reflection asymmetry~\cite{HuSa}
\begin{eqnarray}\label{nbs}
  \frac{n_{L}}{s}\sim \frac{1}{45 T^2_c}\int
  \frac{d\omega}{2\pi}n_0(\omega)[1-n_0(\omega)]\Delta(\omega)
  \Delta \vec p \cdot {\vec v_{\rm w}}\,,
\end{eqnarray}
where $v_{\rm w}\sim0.1$ is the wall velocity and $n_0(\omega) =
1/(e^{\omega/T_c}+1)$ is the Fermi-Dirac distribution. The difference
between $N$ and $L$ momenta for a given energy, $\Delta \vec p\equiv
\vec p_{L}-\vec p_{N}$, is large, due to $O(1)$ mass differences among
$\bar{M}$ and $\bar{M}_N$.  This is welcome because, in the SM,
$\Delta p$ arises only due to the electroweak thermal corrections to
the masses, and is suppressed by $\alpha_W$~\cite{HuSa}.  The
reflection asymmetry $\Delta(\omega)$ is given by~\cite{CKNRev}
$\Delta(\omega)={\rm Tr}\left(R^\dagger_{NL} R_{NL} - \bar
  R^\dagger_{NL}\bar R_{NL}\right)\,,$ where $R_{NL}$ is the
reflection coefficient for $N\to L$ and the bars stand for the CP
conjugated process.
%(A quasi-particle treatment is not necessary.  Unlike in the SM case,
%asymmetry is created without these thermal effects).

\begin{figure}[t]
  \centering
  \includegraphics[width=\columnwidth]{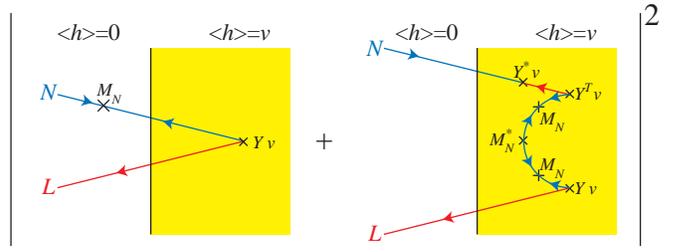}
  \caption{Perturbative calculation of the reflection coefficients
    which picks up the Jarlskog invariant $J_L = \Im m{\rm Tr}[(M_N
    Y)^\dagger (Y^* Y^T M_N M_N^* M_N Y)]$ from the one-particle cuts
    in the amplitudes.}
  \label{fig:reflection}
\end{figure}

In order to calculate the reflection asymmetry from the bubble wall,
we have computed the Green function for our model in the simplifying
limit $\bar{M}_N \gg \bar{M}$.  Using a perturbative expansion in
$Yv/T_c,\,M/\omega$, and for $N_G=2$, we find $\Delta(\omega)\sim
\left(\ell_{T}{ v\over \omega}\right)^{2 N_G} J_L$ where
$\ell_T\sim10/T$ is the mean free path for the leptons~\cite{CKNRev}
(see Fig.~\ref{fig:reflection}).  Below we find that phenomenological
constraints favor somewhat heavier RH Majorana masses, $\bar
M_N\gtrsim \bar Yv$. In this case a suppression in the reflection
asymmetry is expected since the heavy incoming particle is hardly
affected by the potential barrier.  This effect cannot be captured
using an expansion in $M_N/\omega$. We estimated the corresponding
$1/M_N$ suppression by analysing a single generation reflection
problem which can be solved analytically. We indeed found that the
reflection amplitude is further suppressed via $(\bar Yv/\bar M_N)^{2N_G}$.
Consequently, in the relevant region of parameter space the reflection
asymmetry is given by
\begin{eqnarray}
  \Delta(\omega)\sim   \left({\ell_{T} \bar Y^2 v^2\over \omega\bar
      M_N}\right)^{2 N_G} J_L\ \theta(\omega-\bar M_N) 
  \,.\label{Dw} 
\end{eqnarray}  
Below we use the estimate~(\ref{Dw}) even in the case of interest
where $Yv/T_c$ and $M/\omega$ are of order unity.

The rate for baryon production is approximately $dn_{\rm B}/dt\sim
3T_c^2 \Gamma_{\rm Sp} \Delta F$~\cite{CKNRev,HT}. $\Delta F$ is the
free energy difference between two neighboring zero field-strength
configurations, for which $\Delta {\rm B}=\Delta {\rm L}=3$ (provided
that $\Gamma(L\to l \phi)$ is not much slower than other
thermalization rates; see below).  $\Delta F$ is calculated in the
presence of a hypercharge density $n_Y\sim-n_L/2$~\cite{HT}. (More
correctly, one should define a global approximately conserved charge
orthogonal to hypercharge ${\rm B'}={\rm B}-xY$ where in our case
$x=1/7$~\cite{HT}. We verified that this will hardly modify our
results).  In our case we find that $\Delta F \simeq n_{L}/T_c^2$.
The BAU is obtained by integrating $dn_{\rm B}/dt$, which we estimate
via $n_{\rm B}\sim dn_{\rm B}/dt\times \ell_{\rm w}/v_{\rm w}$ where
$\ell_{\rm w}$ is the typical penetration length for the non-zero
global charge which flows through the plasma in the unbroken phase.
With fast massless leptons one expects, from estimation of energy loss
in the plasma~\cite{Eloss}, $T_c \ell_{\rm w}={\cal O}(100)$
~\cite{CKNRev,HT}.  In our case, with semi-relativistic leptons, the
penetration distance is shorter even though the energy loss rate is
very low, because the massive leptons rather quickly lose
their directionality (away from the wall).  For instance, elastic
scattering of a lepton of mass $4T_c$ with a plasmon that carries a
perpendicular momenta, $T_c$, roughly results in 25\% momentum loss in
the original direction of motion for the lepton.  Thus in our case we
estimate $\ell_{\rm w}\sim N_{\rm coll}\ell_{T}$ where $N_{\rm coll}$
is the average number of collisions that a lepton undergoes before its
directionality is lost.  Below, we assume $N_{\rm coll}=2$.

\begin{figure}[!t]
\begin{center}
\includegraphics[width=0.8\columnwidth]{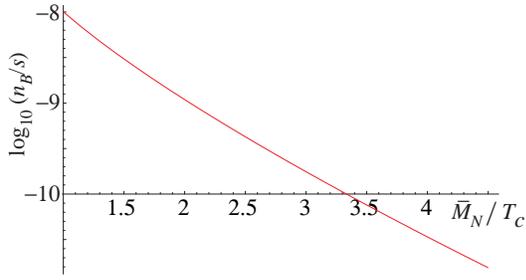}
\caption{$n_{\rm B}/s$ as a function of $\bar M_N/T_c$, for $ \ell_{\rm
    w}\sim 20/T_c,\,N_{\rm G}=2$ and
  $\Delta(\omega)=(T_c/w)^{4}\theta(\omega-\bar M_N)$.
}\label{fignBlab}
\end{center}
\end{figure}

Using Eqs.~(\ref{nbs},\ref{Dw}) we obtain
\begin{eqnarray}
{n_{\rm B}\over s}&\sim& 
3 \Gamma_{\rm Sp} N_{\rm coll} {\ell_{T}\over v_{\rm w}} \,{n_{L}\over 2s}
\,.\label{final}
\end{eqnarray}
Our next step is to identify the dependence of the resultant BAU on
the model fundamental parameters.  Assuming that CP violation is
maximal and taking for simplicity $(\bar Yv)^2 \ell_{T}/T_c\sim 1$ we
find $\Delta(\omega)=(T_c/\omega)^{4}\theta(\omega-\bar M_N)$.  In
addition, fixing $N_{\rm coll}=2$ and $\ell_{T}\sim 10/T_c$ we can
numerically compute the resultant BAU as a function of $\bar M_N/T_c$.
As $\bar{M_N}/T_c$ increases, the BAU rapidly decreases due to both a
Boltzmann suppression factor [see Eq.~(\ref{nbs})] and also a
polynomial one~(\ref{Dw}).  In fig.~\ref{fignBlab} we plot $n_{\rm
  B}/s$ as a function of $\bar M/T_c$.  We find that the observed
asymmetry can be accounted for when
\begin{eqnarray}
  \bar M_N\lesssim 4\times T_c\,.
  \label{Up}\end{eqnarray}
In the Dirac case $N_G=3$ and the resultant
asymmetry is further suppressed.
\vspace*{-0.0cm}

\mysection{Direct Constraints and Tests} In the following we briefly
discuss the direct phenomenological constraints on our scenario and
discuss several ways to directly test our model in the near future.

\noindent{\emph {Electroweak precision measurements}-} Our model requires
additional fields, namely heavy lepton doublets charged under the SM
gauge group. Because they are vector-like, the $S$ parameter is hardly
affected. The Yukawa couplings to the SM Higgs, $Y$, breaks the
custodial isospin symmetry and therefore modify the $T$ parameter.
The additional contribution to the $T$ parameter from the three
vector-like lepton doublets, $T_L$, is similar to a single extra
vector-like top quark~\cite{RS1} $T_{L}\sim T_{t}^{\rm SM}\,{\bar Y^4
  v^4\over \bar m_t^2 M^2}$ where $T_{t}^{\rm SM}\sim1.2$ is the
SM-top contribution to $T$~\cite{PDG} and $m_t$ is the top mass.
Requiring the $T_{L}\lesssim0.2$ we find the following lower bound
\begin{eqnarray}
  {\bar M\over v}\gtrsim 2.5\times \bar Y\,
  \left(\bar Y v\over m_t\right) \,.\label{Down}
\end{eqnarray}
Furthermore, in the Dirac case, the light RH neutrinos are a
linear combination of $N$ and $L^c$, and hence the $Z$ can decay
invisibly to these states.  The rate is proportional to the quartic
power of the corresponding mixing, ${\bar Y^4 v^4\over \bar M^4}$.
Using the 3$\sigma$ range (to allow for at least three
neutrinos)~\cite{EW} we find that 3.01 neutrinos are allowed. For
three extra generations this implies $ \bar M\gtrsim 4\times \bar Y v
\,.$ (This constraint is absent in the Majorana case.)

\noindent{\emph {Lepton flavor violation}-} The Yukawa couplings to the SM
leptons, $y$, are generically not aligned with $M,M_N$ and $Y$ and
hence induce lepton flavor violation.  This will contribute to
processes such as $\mu\to e$ conversion which are highly constrained
by experimental data~\cite{PDG,LFV}.  In our model the SM leptons
couple to the additional fields only through $y$, the Yukawa coupling to
$\phi$.  To avoid these constraints
naturally, a mild hierarchical structure is required for $y$. For
example, in the appropriate basis, $y\sim~{\rm
  diag}\left(10^{-2},10^{-1},10^{-1}\right)\,,$ which is consistent
with the neutrino flavor parameters~\cite{Texture} provided that
$M_N,M,Y$ are anarchical.  The smaller the eigenvalues of $y$, $y_i$,
however, the smaller is the decay rate $L\to l\phi$.  Comparing the decay
rate into SM doublets to the thermalization time scale we find
$\Gamma_L \ell_{\rm w}/v_{\rm w}\sim 10 y^2_i\,\bar M/T_c\sim0.25-1$
which implies further suppression in the resultant BAU (this, given
our crude estimation see fig.~\ref{fignBlab}, is still consistent with
the observed value).  We thus find that there is a tension between
producing sizable BAU and suppressing the contribution to lepton
flavor violation.  Therefore our mechanism typically predicts that the
rates for $\mu\to e$ processes are within the reach of near future
experiments.  In addition, given the above ansatz for $y$, contributions
to neutrinoless double beta decay are evaded~\cite{Gauge}. 

\noindent{\emph {Collider physics}-} From Eqs.~(\ref{Up},\ref{Down}),
our scenario predicts the presence of SU(2) doublets with masses
below the ${\cal O}$(1\,TeV) range. These contain charged particles,
with masses above 100\,GeV from current bounds~\cite{PDG}, that may
be produced and detected at the LHC, or better at the ILC. A clear
signal at the LHC is expected if the vector-like leptons are
sufficiently light, via production of $L^\pm L^0$ which decay to a
single, energetic SM charged lepton. If the Higgs boson is heavier
than the new particles it can also decay invisibly to $N$ and $L$.

Note that the presence of new lepton doublets is crucial to our
mechanism.  One might try to implement our idea replacing the
vector-like doublets with SM singlets. This however would imply that
the light neutrinos are partially sterile which in turn would lead to
non-universality in weak processes.  For instance, the charged-current
universality between $\beta$-decay and $\mu$-decay will be modified.
As these observables are measured below the 0.1\% level~\cite{PDG},
the lower bound on $\bar M$ is strengthen, significantly suppressing
the BAU.  
Alternatively, one might study Type-II seesaw models in which
Majorana neutrino masses are induced by the VEV of an electroweak triplet 
scalar boson; but constraints from the precision electroweak data are 
even more severe.

\mysection{Discussion} In this letter, we showed that the framework of
late neutrino masses naturally leads to a viable model of electroweak
baryogenesis that can be directly tested in near future experiments.

The additional vector-like leptons with mass close to the EW scale
play a crucial role in our scenario. This raises a coincidence
problem since a vector-like mass term is unprotected by the symmetries
of our model and therefore is unrelated to the EW scale, similar to
the $\mu$-problem in the Minimal Supersymmetric SM.  There are however
many proposed solutions to this problem: vector-like mass may be
induced from supersymmetry breaking, additional Higgs fields, or
strong dynamics.

The first-order EW phase transition is mandatory. It may be 
induced, for instance, by higher order terms in the effective potential
that arise from integrating out additional heavy states~\cite{GSW}.

Our baryogenesis mechanism relies only on physics at the TeV scale, 
and hence is compatible with any scheme for new physics above the 
TeV scale. For example there could be flat extra dimensions just above 
the TeV scale, or a warped extra dimension as in the Randall-Sundrum 
scheme; in supersymmetric theories our mechanism allows for a low 
reheating temperature after inflation, solving the various 
moduli/gravitino problems. Finally, our
scenario may accommodate unification by completing the vector-like
leptons to full $SU(5)$ representations.

\vspace*{.02cm}
%%%%%%%%%%%%%%%%%%%%%%%%%%%%%%%%%%%%%%%%%%%%%%%%%%%%%%%%%%%%%%%%%%%%%%%%%%%%
\acknowledgments{We thank K. Agashe, R. Harnik, R. Kitano, Z. Ligeti,
  S. Oliver, M. Papucci, T. Watari \& J. Wells for useful discussions.
  This work was supported in part by the DOE under contracts
  DE-FG02-90ER40542 and DE-AC03-76SF00098 and in part by NSF grant
  PHY-0098840.}

\end{document}